\documentclass[aps,twocolumn,pra,showpacs]{revtex4}

\newcommand{\ket}[1]{\left| #1\right\rangle}

\newcommand{\ketbra}[3][]{\left|#2\right\rangle_{#1}\!\left\langle#3\right|}

\newcommand{\ieac}[0]{{\it et al., }}

\usepackage{amsmath}
\usepackage{amssymb}
\usepackage{dsfont}

\usepackage{times}
\usepackage{graphicx}
\begin{document}


\title{Complete Characterization of Quantum-Optical Processes}


\author{Mirko Lobino,$^{1}$ Dmitry Korystov,$^{1}$ Connor Kupchak,$^{1}$ Eden Figueroa,$^{1}$
Barry C. Sanders,$^{1}$ and A. I. Lvovsky$^{1*}$} \affiliation{$^1$
Institute for Quantum Information Science, University of Calgary,
Calgary, Alberta T2N 1N4, Canada}

\email{lvov@ucalgary.ca}
%
\begin{abstract}
The technologies of quantum information and quantum control are
rapidly improving, but full exploitation of their capabilities
requires complete characterization and assessment of processes that
occur within quantum devices. We present a method for
characterizing, with arbitrarily high accuracy, any quantum optical
process. Our protocol recovers complete knowledge of the process by
studying, via homodyne tomography, its effect on a set of coherent
states, i.e. classical fields produced by common laser sources. We
demonstrate the capability of our protocol by evaluating and
experimentally verifying the effect of a test process on squeezed
vacuum.
\end{abstract}

\maketitle

Construction of a complex machine requires precise characterization
of each component's properties. In electronics, this information is
obtained from network analyzers, which measure circuit response to
simple oscillatory inputs, and reveal the device transfer function.
Optical quantum technologies, which can be used to build quantum
computers \cite{KLM}, precise metrological systems \cite{Kimble},
and unconditionally secure communication \cite{QCrypt}, have similar
characterization requirements. In this context, we are interested in
the process associated with a quantum circuit component, i.e.~in
being able to predict the transformation that an arbitrary quantum
state will undergo when subjected to the action of the component.

A quantum process $\mathcal{E}$ can be represented by a completely
positive, trace-preserving linear map (superoperator) on the linear
space $\mathds{L(H)}$ of all density matrices over Hilbert space
$\mathds{H}$. It can be expressed as a rank-4 tensor that relates
the matrix elements of the output $\mathcal{E}(\hat\rho)$ and input
$\hat\rho$ states in some basis:
\begin{equation}\label{eq.rho_out_final}
    \left[\mathcal{E}(\hat\rho)\right]_{lk}=\sum\limits_{nm}\mathcal{E}_{lk}^{nm}\rho_{nm},
\end{equation}
where summation is from 1 to $\dim\mathds{H}$.

Characterization of a process (known as quantum process tomography,
or QPT) means finding all components of the superoperator tensor. It
can be implemented by determining the output state for each of the
$\left(\dim \mathds{H}\right)^2$ elements of a spanning set of
$\mathds{L(H)}$. Such a direct approach to QPT \cite{Zoller} was
experimentally realized on one qubit teleportation \cite{Laflamme},
the Hamiltonian evolution of vibrational states of atoms in an
optical lattice \cite{Steinberg}, on a two-qubit controlled-NOT gate
\cite{O'Brien,Leung} and Bell-state filter \cite{Steinberg2}.
As an alternative, ancilla-assisted QPT exploits an isomorphism
between processes and states \cite{D'Ariano} and has been used to
characterize a controlled-NOT gate \cite{Blatt} and a general single
qubit gate \cite{O'Brien2,DeMartini}; see \cite{Lidar} for a
comparative review of ancilla-assisted QPT.

Existing QPT suffers from serious shortcomings, including either the
requirement of an unwieldy set of input states for direct QPT or a
high-dimensional entangled input state for ancilla-assisted QPT;
these shortcomings deleteriously affect scalability and restrict
accessible systems to very low dimension.
In optics, QPT has been applied to processes on one and two
dual-rail qubits, with post-selection based on photon coincidences
projecting the input and output states onto these qubit subspaces.
This approach cannot provide complete information about a state or a
process because optical losses, imperfect sources, detector dark
counts, and other imperfections lead to departure from the qubit
subspaces. Post-selected tomography can only estimate the fraction
of such phenomena by comparing  the coincidence rate and the photon
production rate \cite{Steinberg2}.

We introduce a scheme that enables complete characterization of a
general quantum-optical process. We use optical homodyne tomography
followed by maximum likelihood reconstruction to obtain full
information on the process across all photon number sectors and also
the coherence between sectors. The state reconstruction algorithm
provides an efficient method for compensating losses in homodyne
detection \cite{MaxLik}. As inputs, we use only coherent states that
are readily available from a laser source, so our method can be
easily scaled up.

We experimentally test our approach by characterizing a quantum
process that consists of a simultaneous absorption and phase shift.
The reconstructed superoperator allows us to predict, with a
fidelity of over 99\%, the effect of the process on a squeezed
vacuum.

Our method is based on the fact that any density matrix can be
represented as a sum of coherent states' density matrices
\cite{Glauber, Sudarshan}. Although such a representation (the
Glauber-Sudarshan $P$ function) may be highly singular, it can be
arbitrarily closely approximated with a regular $P$ function. By
measuring the process output for many coherent states and exploiting
the linearity, we can predict the process output for any arbitrary
state.

The Glauber-Sudarshan decomposition of a quantum state $\hat\rho$ is
given by
\begin{equation}\label{eq.rho}
    \hat{\rho}=2\int
    P_\rho(\alpha)|\alpha\rangle\langle\alpha|\,d^2\alpha
\end{equation}
and where $P_\rho(\alpha)$ is the state's $P$ function, $\alpha$ is
the coherent state with mean position and momentum observables
$(x,p)=(\sqrt 2{\rm Re}\,\alpha,\sqrt 2{\rm Im}\,\alpha)$. We use
the convention $[\hat x,\hat p]=i$ and integration is performed over
the entire phase space.
Therefore, if we know the effect $|\alpha\rangle\langle
\alpha|\mapsto \hat\varrho(\alpha)=\mathcal E(\ketbra\alpha\alpha)$
of the process on all coherent states, we can predict its effect
upon state $\hat\rho$:
\begin{equation}\label{eq.edec}
   \mathcal E( \hat{\rho})=2\int
    P_\rho(\alpha)\hat \varrho(\alpha)\,d^2\alpha.
\end{equation}

An obstacle to direct application of this approach is posed by
singular behavior of the function $P_\rho(\alpha)$. Indeed, the $P$
function exists only as a generalized function, more singular than
the Dirac delta function, when the corresponding quantum state has
nonclassical features \cite{Leonhardt}.

This can be overcome by applying a theorem proven by Klauder
\cite{Klauder}: for any bounded operator $\hat{\rho}$ there exists
an operator $\hat{\rho}_L$ with continuous and rapidly decreasing
$P$ function arbitrarily close to $\hat{\rho}$ in the trace-class
norm. The Klauder approximation is obtained as follows: we assume
that the Wigner function of $\hat{\rho}$ belongs to the Schwartz
class $\mathcal{S}^2$, i.e. is infinitely smooth and rapidly
decreasing (which is the case for all physically plausible density
matrices). The Fourier transform of the operator's Glauber-Sudarshan
function $P_\rho(\alpha)$ can be expressed as \cite{Leonhardt}
\begin{equation}\label{eq.ptilde}
\tilde{P}_\rho(k_x,k_p)=\tilde{W}_\rho(k_x,k_p)\exp\left(\frac{k_x^2+k_p^2}4\right),
\end{equation}
where $\tilde{W}_\rho(k_x,k_p)$ is the Fourier transform of the
operator's Wigner function. The function defined by Eq.
\ref{eq.ptilde} always exists and is infinitely smooth (albeit not
necessarily square integrable). We multiply
$\tilde{P}_\rho(k_x,k_p)$ by a regularizing function
\begin{equation}\label{eq.func_limit}
    G_L(k_x,k_p)=e^{-\left[f(k_x-L)+f(-k_x-L)+f(k_p-L)+f(-k_p-L)\right]}
\end{equation}
with $f(y)=y^4\exp(-1/y^2)$ for $y>0$, $f(y)=0$ for $y\leq 0$. This
regularizing function is equal to 1 in a square domain of side $2L$
and rapidly drops to zero outside. The product $\tilde
P_{L,\rho}(k_x,k_p)=\tilde P_\rho(k_x,k_p) G(k_x,k_p)$ now belongs
to the Schwartz class. Applying the inverse Fourier transform, we
obtain the new Glauber-Sudarshan function $P_{L,\rho}(\alpha)$,
which defines the Klauder approximation $\hat{\rho}_L$. By choosing
$L$ sufficiently high \cite{SOM}, the operator $\hat{\rho}_L$ can be
made to approximate $\hat\rho$ arbitrarily well (Fig. 4A).

As an example, we applied the Klauder approximation to squeezed
vacuum, a nonclassical state characterized by a highly singular $P$
function whose Fourier transform grows exponentially with increasing
$k_x$ and/or $k_p$. We tested our protocol with a state that has a
noise reduction in the squeezed quadrature of $-1.58$ dB and excess
noise in the orthogonal quadrature of 2.91 dB. The function
$\tilde{P}(k_x,k_p)$ was calculated from the state's density matrix
according to Eq. \ref{eq.ptilde} and subsequently regularized as
described above using $L=5.2$. Fig. 1A shows $\tilde{P}_L(k_x,k_p)$
calculated from our experimental data and Fig.~1B displays its
inverse Fourier transform $P_L(\alpha)$. In Figs.~1C and 1D we
compare the Wigner functions of the original state and the one
obtained from the regularized $P$ function. The two states exhibit a
quantum fidelity of more than 0.9999.
\begin{figure}
 \includegraphics[width=0.5\textwidth]{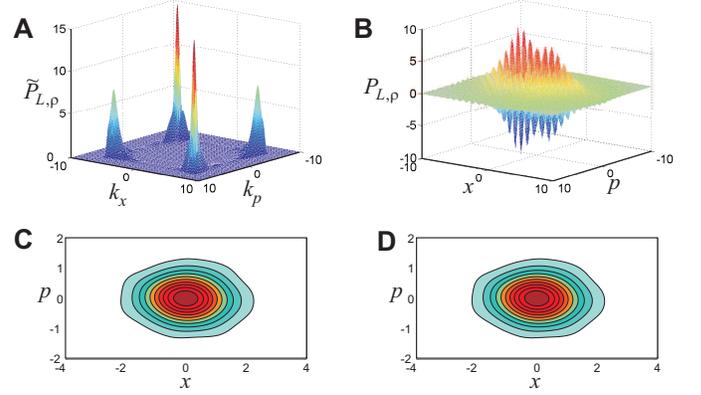}

\caption{Regularized Glauber-Sudarshan decomposition of the squeezed
state. (\textbf{A}) Absolute value of the regularized Fourier
transform of the squeezed vacuum \emph{P} function. (\textbf{B})
Approximated \emph{P} function calculated from the inverse Fourier
transform of $\tilde{P}_{L,\rho}(k_x,k_p)$. (\textbf{C}) and
(\textbf{D}) Wigner representations of, respectively, the measured
and approximated squeezed vacuum states.}
\end{figure}

Although the above method permits finding the process output for an
arbitrary input state, it requires one first to determine the input
state's $P$ function. This step can be avoided by calculating the
process superoperator in the Fock basis, so the output can be found
from the input density matrix according to
Eq.~\ref{eq.rho_out_final}. To this end, we express the
Glauber-Sudarshan function as
\begin{equation}\label{eq.P_Fock}
    P_\rho(\alpha)=\sum\limits_{mn}\rho_{nm}P_{nm}(\alpha),
\end{equation}
where $P_{nm}(\alpha)$ is the $P$ function of the operator $\ketbra
n m$. We now replace these functions by their regularized versions
$P_{L,nm}(\alpha)$ and rewrite Eq. \ref{eq.edec} as
\begin{equation}\label{eq.edec_Fock}
   \mathcal E( \hat{\rho})=2\sum\limits_{nm}\rho_{nm}\int
    P_{L,nm}(\alpha)\hat \varrho(\alpha)\,d^2\alpha,
\end{equation}
from which we determine the process superoperator as
\begin{equation}\label{eq.K_final}
    \mathcal E_{lk}^{nm}=2\int
    P_{L,nm}(\alpha)\varrho_{lk}(\alpha)\,d^2\alpha.
\end{equation}

Prior to applying the latter result to experiments, a number of
practical issues have to be addressed. First, parameter $L$ must be
chosen to ensure proper approximation of input states. The second
issue is that in a realistic experiment, the measurement can be done
only for coherent states whose amplitude does not exceed a certain
maximum $\alpha_{\rm max}$. Finally, the experiment can only be
performed with a finite, discrete set of coherent states. Density
matrix elements $\varrho_{lk}(\alpha)$ for an arbitrary $\alpha$,
required for calculating the superoperator, must then be obtained by
polynomial interpolation. These matters are discussed in \cite{SOM}.

A simplification arises for phase-symmetric processes, in which
there is no phase coherence between the ``processing unit'' and
input states. In this case, if two inputs  $\hat\rho$ and
$\hat\rho_1$ are different by an optical phase shift $\hat
U(\varphi)$, the states $\mathcal E(\hat\rho)$ and $\mathcal
E(\hat\rho_1)$ will differ by the same phase shift:
\begin{equation}\label{eq.shift}
    \mathcal E[\hat U(\varphi)\hat\rho\hat U^\dag(\varphi)]=\hat U(\varphi)\mathcal E(\hat\rho)\hat U^\dag(\varphi).
\end{equation}
Then, if we know the effect of the process on a coherent state
$\ket\alpha$, we also know what happens to $\ket{\alpha
e^{i\varphi}}$, so it is enough to perform measurements on input
coherent states with real, positive amplitudes. For the process
superoperator in the Fock basis, the phase symmetry implies that
$\mathcal E_{mn}^{kl}$ vanishes unless $k-l=m-n$.

The process studied in our experiment was electro-optical amplitude
and phase modulation of the optical field. The process was
implemented using an electro optical modulator (EOM) followed by a
polarizer. The field experienced minimal distortion when a bias
voltage $V_1 = 100$ V was applied to the EOM. Switching the voltage
to $V_2=50$ V produced birefringence, and thus losses at the
polarizer, along with a phase shift.

A continuous-wave Ti:Sapphire laser at 795 nm was the coherent state
source used for the device characterization. We reconstructed the
input and output states at 11 different input amplitude levels
between $\alpha_1=0$ and $\alpha_{11}=10.9$. In order to keep track
of the relative phase shift, the EOM voltage was switched between
$V_1$ and $V_2$ every 100 $\mu$s (Fig.~2A, top) while the phase of
the local oscillator was linearly scanned by a piezoelectric
transducer at 100 Hz. The homodyne photocurrent was recorded with an
oscilloscope. To obtain quadrature measurements, the photocurrent
was integrated over time intervals of 20 ns. The bottom plot in Fig.
2A shows the recovered quadrature values after normalization to the
vacuum noise. The time dependence of the local oscillator phase was
recovered from the slow, sinusoidal variation of the average
homodyne photocurrent as a function of time.

In this manner, for each input amplitude, we sampled 50,000 phase
and quadrature values for both the input and output states and used
them to calculate density matrices by likelihood maximization
\cite{MaxLik,MaxLik2} (Fig.~2B). The output state reconstruction
showed a phase shift of $36^\circ$ and a loss of 34 \% with respect
to the input state.

\begin{figure}
 \includegraphics[width=0.5\textwidth]{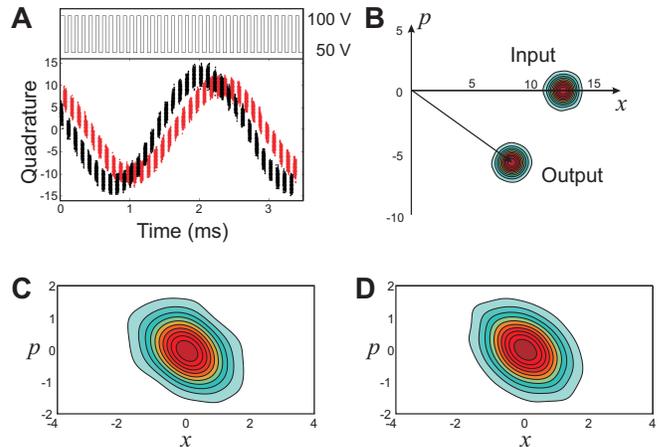}

\caption{(\textbf{A}) Time-dependent quadrature values acquired from
homodyne detection of a coherent state with input $\alpha=8.3$.
Black dots correspond to the state before the process; red dots,
after the process. The top curve shows the EOM driving voltage.
(\textbf{B}) The Wigner function of the coherent state before and
after the process. (\textbf{C}) and (\textbf{D}) Wigner
representations of the measured output squeezed state compared to
the one obtained from process tomography.}
\end{figure}

The interpolated experimental data have been used to determine the
process superoperator tensor. We used the phase symmetry assumption
in Eq. \ref{eq.shift}, which is justified by the fact that the EOM
driver is independent from the master laser. The elements
$\mathcal E_{kk}^{mm}$ of the tensor in the photon number basis are
plotted in Fig.~3A. This plot should be interpreted as follows: for
a given input Fock state $\ket m$, the values of $\mathcal
E_{kk}^{mm}$ give the diagonal elements of the output density
matrix. For example, the single-photon state $\ket 1$ after passing
through the EOM will be transformed into a statistical mixture of
the single-photon and vacuum states.
A theoretical prediction for the process tensor has been calculated
using the Bernoulli transformation to account for a lossy channel
and a phase shift superoperator; the superoperator diagonal elements
in the Fock basis are displayed in Fig. 3B, and these diagonal
elements bear close resemblance to the experimental result. A
similar agreement was also obtained for non-diagonal terms of the
superoperator, but it is not shown here.

\begin{figure}
 \includegraphics[width=0.3\textwidth]{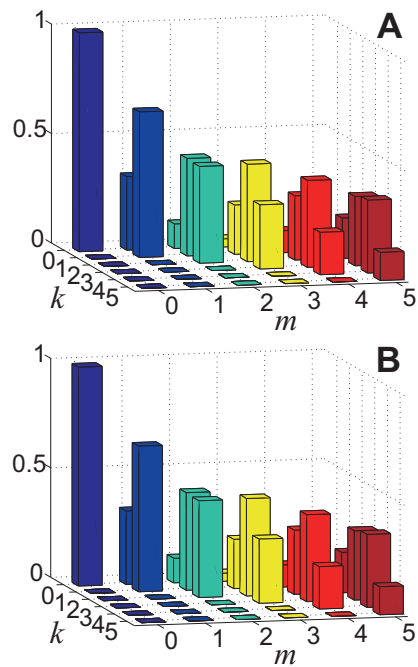}
\caption{The ``diagonal'' values of the superoperator $\mathcal
E_{kk}^{mm}$. (\textbf{A}): as obtained in the experiment.
(\textbf{B}): theoretical model.}
\end{figure}

For additional verification, we applied this result to predict the
effect of the device on the squeezed vacuum state described in the
previous section. This state was produced by pumping an optical
parametric amplifier (OPA) in bow tie configuration with the second
harmonic of the Ti:Sapphire laser and using a periodically poled
KTiOPO$_4$ crystal as nonlinear medium\cite{OPA,QMSL}.

The state before (Fig.~1C) and after (Fig.~2C) the process was
reconstructed using homodyne tomography as described above. By
applying the process superoperator to the input squeezed state, we
predict the process output (Fig. 2D). The maximum (minimum)
quadrature noise variance amounted to 2.19($-$1.07) dB for the
measured state, and 2.15($-$0.95) dB for the prediction,
corresponding to a quantum fidelity of 0.9935$\pm$0.0002.

Whereas here we demonstrate our tomographic method for single-mode
inputs, multimode or multichannel processes can be characterized
using multimode P representation, multiple homodyne detectors and
feeding product coherent states as inputs. Theoretical and
experimental analysis of the multi-mode setting will be discussed
elsewhere.

Our method overcomes significant limitations of previous optical QPT
schemes. Process characterization is not restricted to a Hilbert
space associated with a specific qubit, and thus reveals the
imperfections of a quantum information processing unit.
Additionally, it uses only coherent states as inputs, which are
readily available from the laser and whose intensities and phases
are easily manipulated. This permits characterization of complex
processes employed in quantum information processing and
communication.
\\

\appendix{\textbf{Supporting Material.}} In order to compute
the P function $P_{L,\rho}(\alpha)$ by the Klauder approximation
\cite{Klauder} and the superoperator terms from
\begin{equation}\label{eq.K_final}
    \mathcal E_{lk}^{nm}=2\int
    P_{L,nm}(\alpha)\varrho_{lk}(\alpha)\,d^2\alpha,
\end{equation}
where $P_{L,nm}(\alpha)$ is the regularized P function of the
operator $\ketbra n m$, the values of two parameters must be set:
the size $L$ of the Fourier domain, and the maximum coherent state
amplitude $\alpha_{\rm max}$. As evidenced by Fig.~4A, no value of
$L$ can ensure universally high fidelity for all optical states.
However, for many practical applications it is reasonable to
restrict the input states to subspace $\mathds{H}_N={\rm span}(\ket
0, \ldots, \ket N)$ with a limited number of photons. Under this
restriction, we can seek the lowest $L$ that, in the worst case over
all density matrices in~$\mathds{H}_N$, yields an approximation to
the input state with required fidelity~$F$.

We employed the genetic optimization algorithm \cite{Chambers} to
calculate the required value of $L$ for several low values of $N$
and ascertained that the worst case scenario corresponds to the Fock
state $\ket N$. This is not surprising, because the $P$ function of
this state contains the highest order derivative of the Dirac delta
in $\mathds{H}_N$. We assume this rule to hold for arbitrary $N$.

Restricting the amplitude of coherent states for which the
measurements are performed effectively introduces finite integration
limits in Eq. 1, entailing additional fidelity loss. Fig.~4B shows
the lowest $\alpha_{\rm max}$ required for approximating the
$N$-photon Fock state (which, again, appears to be the worst case in
$\mathds{H}_N$) with a 99 per cent fidelity. Notably, this quantity
is much larger than that expected from the behavior of the
corresponding Wigner function. This is because the regularized $P$
function exhibits strong oscillations in a much wider phase space
region than does the Wigner function.

Even with a finite $\alpha_{\rm max}$, reconstructing the
superoperator requires knowing the process output for a continuum of
coherent states. In an experiment, we choose a finite set of inputs
$\ket{\alpha_i}$ from a laser source with varying, but known,
amplitudes and phases, and perform homodyne tomography to determine
the corresponding output density matrices $\varrho_{lk}(\alpha_i)$
in the Fock basis. For an arbitrary $\alpha$, density matrix
elements $\varrho_{lk}(\alpha)$ are then obtained by polynomial
interpolation.

The degree of the polynomial over~$\alpha$ can be reduced for
Gaussian and near-Gaussian maps, which comprise all linear-optical
processes and $\chi^{(2)}$ nonlinearities. For each measured output
$\hat\varrho(\alpha_i)$ we infer its mean complex amplitude
$\langle\hat{a}\rangle=\textrm{Tr}\left(\hat{a}\varrho(\alpha_i)\right)$.
Then we apply the displacement operator
    \begin{equation}
        \varrho_i^{\langle\hat{a}\rangle}=D(-\langle\hat{a}\rangle)
            \varrho(\alpha_i)D^\dagger(-\langle\hat{a}\rangle).
    \end{equation}
 to center the Wigner function at the origin of the phase space.
The matrix elements $\varrho_{lk}^{\langle\hat{a}\rangle}$, as well
as the mean amplitude $\langle\hat{a}\rangle$, can now be fitted
with a low-degree polynomial over $\alpha$, allowing for a highly
efficient reconstruction process. Good performance of this algorithm
for Gaussian preserving maps is explained by the fact that Gaussian
states are entirely characterized by the first and second centered
moments $\overline{(x-\sqrt 2\,{\rm
Re}\langle\hat{a}\rangle)^m(p-\sqrt 2\,{\rm
Im}\langle\hat{a}\rangle)^n}$ of their Wigner distribution.
\begin{figure}[h!]
 \includegraphics[width=0.4\textwidth]{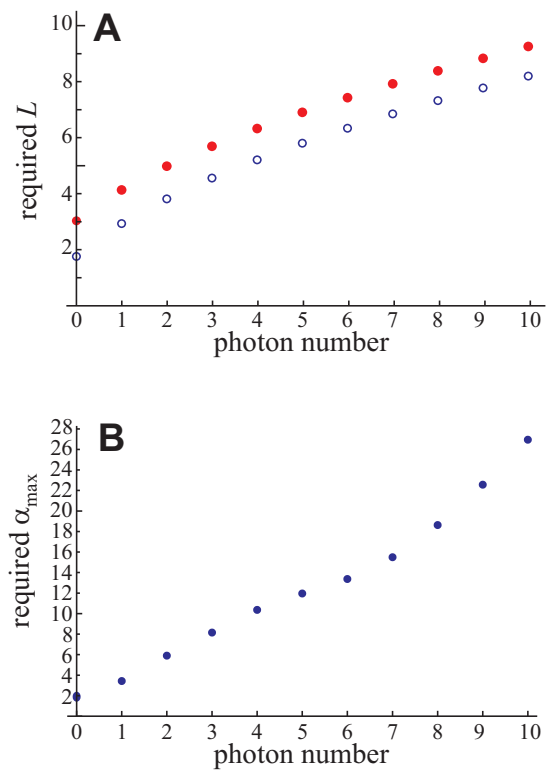}
\caption{Requirements for faithful approximation of photon number
states with a regularized, restricted Glauber-Sudarshan
decomposition. (\textbf{A}): the low pass filtering parameter $L$
needed to achieve a 99.99-\% fidelity ($\bullet$) and 99-\% fidelity
($\circ$). (\textbf{B}): the lowest $\alpha_{max}$ required for a
99-\% fidelity.}
\end{figure}

\end{document}